\documentclass{JHEP}
\usepackage{epsfig}
\usepackage{amssymb}
\usepackage[active]{srcltx}

\newcommand{\bea}{\begin{eqnarray}}
\newcommand{\eea}{\end{eqnarray}}
\newcommand{\bean}{\begin{eqnarray*}}
\newcommand{\eean}{\end{eqnarray*}}

\def\O #1{\overline{#1}}
\def\wt #1{\widetilde{#1}}

\def\det{\mathop{\rm det}}

\def\da{{\rm d}}
\def\d{\partial}

\def\th{{\theta}}
\def\bth{{\overline{\theta}}}
\def\a{{\alpha}}
\def\da{{\dot\alpha}}

\def\b{{\beta}}

\def\la{\lambda}

\def\A{\varphi}

\preprint{
SNUST 030702\\
{\tt hep-th/0307275}}
\title{Renormalizability of
Non(anti)commutative Gauge Theories with ${\cal N}={1 \over 2}$
Supersymmetry \footnote{This work was supported in part by NSF
grant PHY-0070928 (OL), the KOSEF Interdisciplinary Research Grant
98-07-02-07-01-5 (SJR), and the KOSEF Leading Scientist Grant
(SJR).}}
\author{Oleg Lunin ${}^a$ , \,  Soo-Jong Rey ${}^{a,b}$\\
~~~~~~~~~~~~~~\\
${}^a$ Institute for Advanced Study\\
 Einstein Drive, Princeton NJ 08540 USA\\
 ~~~~~~~~~~~~~~~~\\
${}^b$ School of Physics \& BK-21 Physics Division\\
Seoul National University, Seoul 151-747 KOREA\\
~~~~~~~~~~~~~~~~~~~\\
 \email{lunin@ias.edu \quad sjrey@gravity.snu.ac.kr} }
\abstract{Non(anti)commutative gauge theories are supersymmetric
Yang-Mills and matter system defined on a deformed superspace
whose coordinates obey non(anti)commutative algebra. We prove that
these theories in four dimensions with ${\cal N}={1 \over 2}$
supersymmetry are renormalizable to all orders in perturbation
theory. Our proof is based on operator analysis and symmetry
arguments. In a case when the Grassman-even coordinates are
commutative, deformation induced by non(anti)commutativity of the
Grassman-odd coordinates contains operators of dimension-four or
higher. Nevertheless, they do not lead to power divergences in a
loop diagram because of absence of operators Hermitian-conjugate
to them. In a case when the Grassman-even coordinates are
noncommutative, the ultraviolet-infrared mixing makes the theory
renormalizable by the planar diagrams, and the deformed operators
are not renormalized at all. We also elucidate relation at quantum
level between non(anti)commutative deformation and ${\cal N}={1
\over 2}$ supersymmetry. We point out that the star product
structure dictates a specific relation for renormalization among
the deformed operators.}
\keywords{superstring, gauge theory, noncommutative geometry}
\begin{document}
%%%%%%%%%%%%%%%%%%%%%%%%%%%%%%%%%%%%%%%%%%%%%%%%%%%%%%%%%%%%%%%%%
\section{Introduction}
%%%%%%%%%%%%%%%%%%%%%%%%%%%%%%%%%%%%%%%%%%%%%%%%%%%%%%%%%%%%%%%%%
Deformation of ordinary superspace has attracted renewed interest,
sparked off in part by the study of string dynamics in the
background of Ramond-Ramond flux \cite{berkovits}. A situation
where this sort of issue is prominently brought up is string
theoretic understanding of gauge theory - matrix model
correspondence put forward by Dijkgraaf and Vafa \cite{dv} (See
also \cite{misc} for other closely related motivations and
interesting applications.).

Consider a D-brane in the background of Ramond-Ramond flux and
open string dynamics on it. Much the same way the Kalb-Ramond
2-form potential $B_{mn}$  affects algebra obeyed by Grassman-even
coordinates \cite{douglas}, the Ramond-Ramond flux does so for the
algebra of Grassman-odd coordinates
\cite{deboer,oogurivafa,seiberg,berkovitsseiberg}. For example,
take a (space-filling) D3-brane of the Type IIB superstring theory
compactified on a Calabi-Yau 3-fold $X$, where the Kalb-Ramond
2-form potential and the self-dual graviphoton flux are turned on
along the flat spacetime, $\mathbb{R}^4$. If they are constant
throughout $\mathbb{R}^4$, these background fields do not produce
energy-momentum tensor and hence do not back-react to the
geometry. Nevertheless, these background fields render nontrivial
effect on the D3-brane worldvolume
\cite{seiberg,berkovitsseiberg}: ${\cal N}=1$ supersymmetry is
deformed to ${\cal N}={1 \over 2}$ supersymmetry. In terms of
chiral coordinates $z \equiv (y^m, \th^\a, \bth^\da)$
$(m=1,\cdots,4, \a, \da = 1,2)$ of the underlying ${\cal N}=1$
superspace, the deformed superspace is such that
\bea \left[ y^m, y^n \right] = i \Theta^{mn} \qquad \mbox{and}
\qquad \left\{ \th^\a, \th^\b \right\} = C^{\a\b}, \label{nac}
\eea
where $\Theta^{[mn]}$, $C^{(\a\b)}$ refer to combinations of
background NS-NS (Neveu-Schwarz - Neveu-Schwarz) and R-R
(Ramond-Ramond) fields. In a suitable low-energy decoupling limit
($\ell_{\rm st} \rightarrow 0$ and rescaling of the background
fields), the open string dynamics on D3-brane worldvolume is
described by Yang-Mills fields. With the background fields turned
on, underlying ${\cal N}=1$ superspace is deformed to ${\cal N}={1
\over 2}$ superspace as in (\ref{nac}). Accordingly, underlying
${\cal N}=1$ supersymmetric gauge theory is deformed to ${\cal
N}={1 \over 2}$ supersymmetric gauge theory, in which the ordinary
product between (super)fields is replaced by the star product
\bea \star = \exp \left( {i \over 2} \Theta^{mn}
\overleftarrow{\partial \over \partial
y^m}\overrightarrow{\partial \over \partial y^n} - {1 \over 2}
C^{\a\b} \overleftarrow{\partial \over \partial \th^\a}
\overrightarrow{\partial \over \partial \th^\b} \right).
\label{star} \eea
Thus, study of {\sl open} string dynamics in the background
Ramond-Ramond flux calls first for thorough understanding of the
deformed supersymmetric gauge theory.

In this work, we take a step toward this direction
\footnote{Quantum dynamics of non(anti)commutative Wess-Zumino
model was studied recently in \cite{brittofengrey1, jungtay,
brittofengrey2, grisaru}.}. We study quantum dynamics of ${\cal
N}={1 \over 2}$ supersymmetric gauge theory and prove that the
theory is renormalizable to all orders in perturbation theory. We
also offer a deeper understanding concerning relations between the
non(anti)commutative deformation and the ${\cal N}={1 \over 2}$
supersymmetry. We show that the non(anti)commutative deformation
of ${\cal N}=1$ supersymmetric gauge theory gives rise to ${\cal
N}={1 \over 2}$ one with a specific choice of coefficients for
various deformed terms in the Lagrangian. The choice is dictated
by the star product (\ref{star}). On the other hand, generic
${\cal N}={1 \over 2}$ supersymmetric gauge theories permit
arbitrary coefficients and do not automatically bear the
structure of the star
product (\ref{star}). Nevertheless, for arbitrary
coefficients, we show that the theory is renormalizable.

In a related context, it was found recently that ${\cal N}={1\over
2}$ Wess-Zumino model is renormalizable \cite{grisaru}. The model,
however, was not the one obtained by non(anti)commutative
deformation of underlying ${\cal N}=1$ Wess-Zumino model, but the
one in which several operators (breaking the ${\cal N}=1$
supersymmetry to ${\cal N}={1 \over 2}$) were added by hand. In
contrast, ${\cal N}={1 \over 2}$ supersymmetric gauge theory
obtained via non(anti)commutative deformation is renormalizable by
itself, and there is no need to add new operators as in the
Wess-Zumino model case. This pleasant surprise arises because of
several (pseudo)symmetries underlying the theory. More
specifically, helicity conservation, R-symmetry, and flavor
symmetries constrain possible ultraviolet-divergent operators and
counterterms.

It is interesting that the theory is renormalizable even though it
contains operators ${\cal O}$ of mass-dimension five or higher.
Typically, a quantum field theory is defined by a hermitian
self-conjugate Lagrangian. In this case, one can show that
insertion of such operators into a loop diagram renders the
diagram power-divergent (thus the theory becomes
nonrenormalizable), one can always connect an operator ${\cal O}$
with some diagrams, for example, by starting with a pair ${\cal
O}{\cal O}^\dagger$ and using the vertices present in the theory.
This constitutes the crux of power-counting in ordinary quantum
field theories. On the other hand, in ${\cal N}=\frac{1}{2}$
theories obtained by the non(anti)commutative deformation
(\ref{nac}), the hermiticity is broken as one deforms the chiral
Grassman-odd coordinates $\th^\a$ but not the antichiral ones
$\bth^\da$. Therefore, higher-dimension operators are not
accompanied by its hermitian-conjugates, and the ordinary
power-counting argument would not apply. We will comment more on
relation between lack of hermiticity and power-counting
renormalizability in section 3.

This work is organized as follows. In section 2, we set up the
${\cal N}={1 \over 2}$ supersymmetric gauge theory with
multi-flavor matter. We take the gauge group $G=$ U(N). In section
3, by power-counting and symmetry arguments, we show that the
theory is renormalizable. In section 4, we present operator
analysis and classify requisite counterterms. We show that the
theory contains only a finite number of operators receiving
logarithmic divergences. In section 5, we consider turning on
noncommutativity $\Theta^{ab}$ for Grassman-even coordinates, as
in (\ref{nac}). Making use of known features concerning UV-IR
(ultraviolet-infrared) mixing \cite{uvir} (see \cite{sjrey} for a
coherent account for the phenomenon), we show that the gauge
theories are again renormalizable, but in a different sense that
involves only the planar diagrams. We also discuss interplay
between star product (\ref{star}) and ${\cal N}={1 \over 2}$
supersymmetry. We assert that, in order for the quantum theory to
retain the star product structure, radiative corrections for
various deformed operators ought to obey certain scaling
relations. We conclude in section 6 with discussions on issues
worthy for further investigation.
%%%%%%%%%%%%%%%%%%%%%%%%%%%%%%%%%%%%%%%%%%%%%%%%%%%%%%%%%%%%%%%%%
\section{{\cal N}$={1 \over 2} $ super Yang-Mills theory}
%%%%%%%%%%%%%%%%%%%%%%%%%%%%%%%%%%%%%%%%%%%%%%%%%%%%%%%%%%%%%%%%%
The ${\cal N}={1 \over 2}$ supersymmetric gauge theory is defined
as follows \cite{seiberg}. Start with real vector superfield $V$,
valued in Lie algebra $g=$u(N), and define non(anti)commutative
gauge transformation
\bea e^V_\star \rightarrow e^{V'}_\star = e^{- i \O \Lambda}_\star
\star e^V_\star \star e^{+i \Lambda}_\star \label{gt} \eea
where the gauge transformation functions $\Lambda, \O \Lambda$ are
chiral and antichiral superfields.  The chiral and antichiral
field-strength superfields are defined by \footnote{We follow the
convention of Wess and Bagger \cite{wessbagger}, but rescale the
vector superfield as $V_{\rm ours} \leftrightarrow (2 g) \, V_{\rm
WB}$. Also we recall the relation between the conventional coupling
constant $g$ and the coupling constant $e_{\rm WB}$ of \cite{wessbagger}:
$e_{\rm WB}=2g$.}
\bea W_\a &\equiv& - {1 \over 4} \O D \O D \Big( e^{-V}_\star
\star D_\a e^V_\star \Big) \nonumber \\
\O W_\da &\equiv& + {1 \over 4} D D \Big( e^V_\star \star \O D_\da
e^{-V}_\star \Big). \nonumber\eea
Under the non(anti)commutative gauge transformations
(\ref{gt}), these field-strength superfields transform as
\bea W_\a &\rightarrow& W_\a' := e^{- i \Lambda}_\star
\star W_\a \star e^{+i \Lambda}, \nonumber \\
\O W_\da &\rightarrow& \O W_\da' := e^{- i \O \Lambda}_\star \star
\O W_\da \star e^{+i \O \Lambda}_\star. \nonumber \eea
Fixing the gauge freedom by non(anti)commutative counterpart of
the Wess-Zumino gauge, the vector superfield is reduced to
\bea V(y, \th, \bth) &=& - \th \sigma^m \bth A_m(y) + i \th\th\bth
\O \lambda(y) - i \bth\bth\th^\a \Big(\lambda(y) + {1 \over 4}
\epsilon_{\a\b} C^{\b\gamma} \sigma^m_{\gamma \dot\gamma} \{ \O
\lambda^{\dot\gamma} , A_m \} \Big) \nonumber \\
&+& {1 \over 2} \th\th \bth\bth \Big( D(y) - i \partial_m A^m (y)
\Big), \nonumber \eea
where, following \cite{seiberg}, $\bth\bth\th$-term is modified so
that the standard gauge transformation rule follows for component
fields.

We also couple matter system by introducing a set of chiral
superfields transforming in appropriate representations of the
gauge group $G$. For example, a matter of $N_F$ flavors with
vectorlike coupling is described by $\Phi_f, \wt\Phi_{\tilde f}$
($f, \wt f = 1, \cdots, N_F$) transforming in a pair of conjugate
representations ${\tt R}$ and $\O {\tt R}$:
\bea \Phi_f(y, \th) &=& \A_f(y) + \sqrt{2} \th \psi_f (y) + \th\th
F_f(y) \nonumber \\
\wt \Phi_{\wt f} (y, \th) &=& \wt \A_{\wt f} (y) + \sqrt{2} \th
\wt \psi_{\wt f} (y) + \th\th \wt F_{\wt f} (y) \nonumber \\
\O \Phi_f (\O y, \bth) &=& \O \A_f (\O y) + \sqrt{2} \bth \O
\psi_f(\O y) + \bth\bth \Big( \O F_f(\O y) + i C^{mn} \partial_m
(\O \A_f A_n)(\O y) -
{1 \over 4} C^{mn} \O \A_f A_m A_n (\O y) \Big) \nonumber \\
\O {\wt \Phi}_{\wt f} (\O y, \bth) &=& \O {\wt \A}_{\wt f} (\O y)
+ \sqrt{2} \bth \O {\wt \psi}_{\wt f} (\O y) + \bth\bth \Big( \O
{\wt F}_{\wt f} (\O y) + i C^{mn} \partial_m (\O {\wt \A}_{\wt f}
A_n)(\O y) - {1 \over 4} C^{mn} \O {\wt \A}_{\wt f} A_m A_n (\O y)
\Big) \nonumber \eea
where $\O y^m := y^m - 2 i \th \sigma^m \bth$. Again, to ensure
the standard gauge transformations of component fields, we have
modified the $\bth\bth$-term \cite{ito}. Then, the ${\cal N}={1
\over 2}$ supersymmetric gauge theory is described by the
Lagrangian with gauge coupling parameter $\tau = {\theta \over 2
\pi} + i {4 \pi \over g^2}$:
\bea\label{LagrSuper} L &=& \left[\O \Phi_f \star e^{V}_\star
\star \Phi_f + \O {\wt \Phi}_{\tilde f}
\star e^{-V}_\star \star \wt \Phi_{\tilde f}
\right]_{\th\th\bth\bth} + W_\star(\Phi_f, \wt \Phi_{\tilde f})
\Big|_{\th\th}
+
\O W_\star(\O \Phi_f, \O{\wt \Phi}_{\tilde f})
\Big|_{\bth\bth}\nonumber \\
&+& {\rm Tr} \Big[{i \tau \over 16 \pi} W^\a \star W_\a
\Big]_{\th\th} - {\rm Tr} \Big[ {i \O \tau \over 16 \pi} \O W^\da
\star \O W_\da \Big]_{\bth\bth}. \eea

We are mostly concerned with features arising from
non(anti)commutativity of Grassman-odd coordinates, so we will set
$\Theta^{mn} = 0$ in what follows. For simplicity, we will also
set the coupling parameter $\tau =i { 4\pi \over g^2}$. Upon
expanding the star product among the fields explicitly, the
Lagrangian (\ref{LagrSuper}) is decomposable into a sum of an
ordinary part identical to the ${\cal N}=1$ supersymmetric gauge
theory:
\bea L_{{\cal N}=1} = \left[\O \Phi_f e^{V} \Phi_f + \O {\wt
\Phi}_{\wt f} e^{-V} \wt \Phi_{\wt f} \right]_{\th\th\bth\bth} +
\left[ W(\Phi_f, \wt \Phi_{\wt f}) + {\rm Tr} {i \tau \over 16 \pi
} W^\a W_\a \right]_{\th\th} + \mbox{(h.c.)}, \nonumber \eea
and a deformation part which depends on powers of the
non(anti)commutativity parameter $C^{\a\b}$. This deformation can
be expressed as D-terms involving spurion fields \cite{spurion}:
\bea L_{{\cal N}={1\over2}} &=& \Big[ {1 \over 4 g^2} {\rm Tr}
U^{\a\b} (D_\a W_\b) \O W^\da \O W_\da - { U^{\a\b} \over 2
\sqrt{2}} \Big\{ (D_\a \O D_\da \O \Phi_f) \O W^\da D_\b \Phi_f +
(D_\a \O D_\da \O {\wt\Phi}_{\wt f}) \O W^\da D_\b
{\wt\Phi}_{\wt f} \Big\} \nonumber \\
&&- {1 \over 2} U^{\a\b} \Big\{ \O \Phi_f D_\a W_\b D^2 \Phi_f +
\O {\wt \Phi}_{\wt f} D_\a W_\b \wt \Phi_{\wt f} \Big\} \nonumber
\\
&&+ {U \over 16 g^2} {\rm Tr} (\O W^\da \O W_\da)^2 + {U \over 16}
\Big\{ \O \Phi_f \O W_\da^2 D^2 \Phi_f + \O {\wt\Phi}_{\wt f} \O
W_\da^2 D^2 {\wt\Phi}_{\wt f} \Big\}
\Big]_{\th\th\bth\bth},\nonumber \eea
where the spurion fields are \footnote{We denote $|C|^2 \equiv 4
\det (C^{\a\b}) = C^{mn} C_{mn}$.}
\bea U^{\a\b} = C^{\a\b} \th\th\bth\bth \qquad \mbox{and} \qquad U
= |\, C|^2 \th\th\bth\bth. \nonumber \eea

For $W(\Phi, \wt \Phi) = \O W(\O \Phi, \O {\wt \Phi}) = 0$, which
we are mostly concerned with in this work, the matter fields are
massless. In this case, the theory has a chiral flavor symmetry:
\bea\label{flavor} {\rm U}_{\rm A}(1): \qquad \Phi_f\rightarrow
e^{i\gamma}\Phi_f,\qquad
{\wt\Phi}_{\tilde f}\rightarrow e^{i\gamma}{\wt\Phi}_{\tilde f}.
\eea
In the next section, this symmetry will play a useful role for
operator analysis and proof of renormalizability.

Under the ${\cal N}={1 \over 2}$ supersymmetry, component fields
transform as
\bea \delta A_m &=& - i \O \lambda \O \sigma_m \epsilon \nonumber \\
\delta \lambda_\a &=& i \epsilon_\a D + (\sigma^{mn} \epsilon)_\a
\Big[ F_{mn} + { i \over 2} C_{mn} \O \lambda \O \lambda \Big];
\qquad \delta \O \lambda_\da = 0 \nonumber \\
\delta D &=& - \epsilon \sigma^m \nabla_m \O \lambda
\label{susytransfv}\eea
for the vector superfield, as
\bea \delta \A_f &=& \sqrt{2} \epsilon \psi_f \qquad \qquad \delta
\O \A_f = 0 \nonumber \\
\delta \psi^\a_f &=& \sqrt{2} \epsilon^\a F_f \quad \qquad \delta
\O \psi_{\da \, f} = - i \sqrt{2} (\nabla_m \O \A_f) (\epsilon
\sigma^m)_\da \nonumber \\
\delta F_f &=& 0 \qquad\qquad \delta \O F_f = - i \sqrt{2}
\nabla_m \O \psi_f \O \sigma^m \epsilon - i \O \A_f \epsilon
\lambda + C^{mn} \nabla_m (\O \A_f \epsilon \sigma_n \O \lambda)
\label{susytransfm} \eea
for the matter superfield $\Phi_f, \O \Phi_{f}$, and similarly for
$\wt \Phi_{\tilde f}, \O {\wt \Phi}_{\tilde f}$.

The Lagrangian (\ref{LagrSuper}) is given in terms of component
fields \cite{seiberg,ito} as \footnote{The Lagrangian is
consistent with the normalization of Wess and Bagger after the
vector superfield is rescaled as $V_{\rm ours} = (2 g) V_{\rm
WB}$.} :
\bea\label{SQCD} L&=&\frac{1}{2g^2}\mbox{Tr}\left\{- {1 \over 4}
F_{mn}F^{mn}- i{\O \la}{\O \sigma}^m \nabla_m \la
+ \frac{1}{2}D^2\right\}\nonumber\\
&+&{\O F}_fF_f-i{\O \psi}_f{\O \sigma}^m \nabla_m \psi_f- \nabla_m
{\O \A}_f\nabla_m {\A}_f+\frac{1}{2}{\O \A}_f D{\A}_f+
\frac{i}{\sqrt{2}}({\O \A}_f{\la}\psi_f-{\O \psi}_f{\O \la}A_f)
\\
&-&\frac{i}{4g^2}C^{mn}\mbox{Tr}\left\{F_{mn}{\O \la}{\O
\la}\right\}+ \frac{|C|^2}{16 g^2}\mbox{Tr}({\O \la}{\O \la})^2-
\frac{1}{\sqrt{2}}C^{\alpha\beta} (\nabla_m {\O \A}_f)
\sigma^m_{\alpha{\dot\alpha}} {\O \la}^{\dot\alpha}\psi_{\beta \,
f} \\
&+&\frac{i}{2}C^{mn}{\O \A}_f F_{mn}F_f + \frac{|C|^2}{16}{\O
\A}_f{\O \la}{\O \la}F_f \nonumber.\nonumber\eea
To shorten the expressions, we suppressed terms involving the
superfields ${\wt\Phi}, \O {\wt \Phi}$ in the above Lagrangian.
These terms are obtainable by replacing $\Phi_f, \O \Phi_f$ in
(\ref{SQCD}) to $\wt \Phi_{\tilde f}, \O {\wt \Phi}_{\tilde f}$.

Finally, in the normalization we adopt, the covariant derivative
and field strength are defined by
\bea \nabla_m \A=\d_m A+\frac{i}{2}A_m \A,\qquad \nabla_m \la=\d_m
\la+\frac{i}{2}[A_m,\la],\qquad F_{mn}=\d_m
A_n-\d_nA_m+\frac{i}{2}[A_m,A_n]. \nonumber \eea
In the next section, we will prove that ${\cal N}={1 \over 2}$
supersymmetric gauge theory described by (\ref{SQCD}) is
renormalizable to all orders in perturbation theory.

\EPSFIGURE[ht]{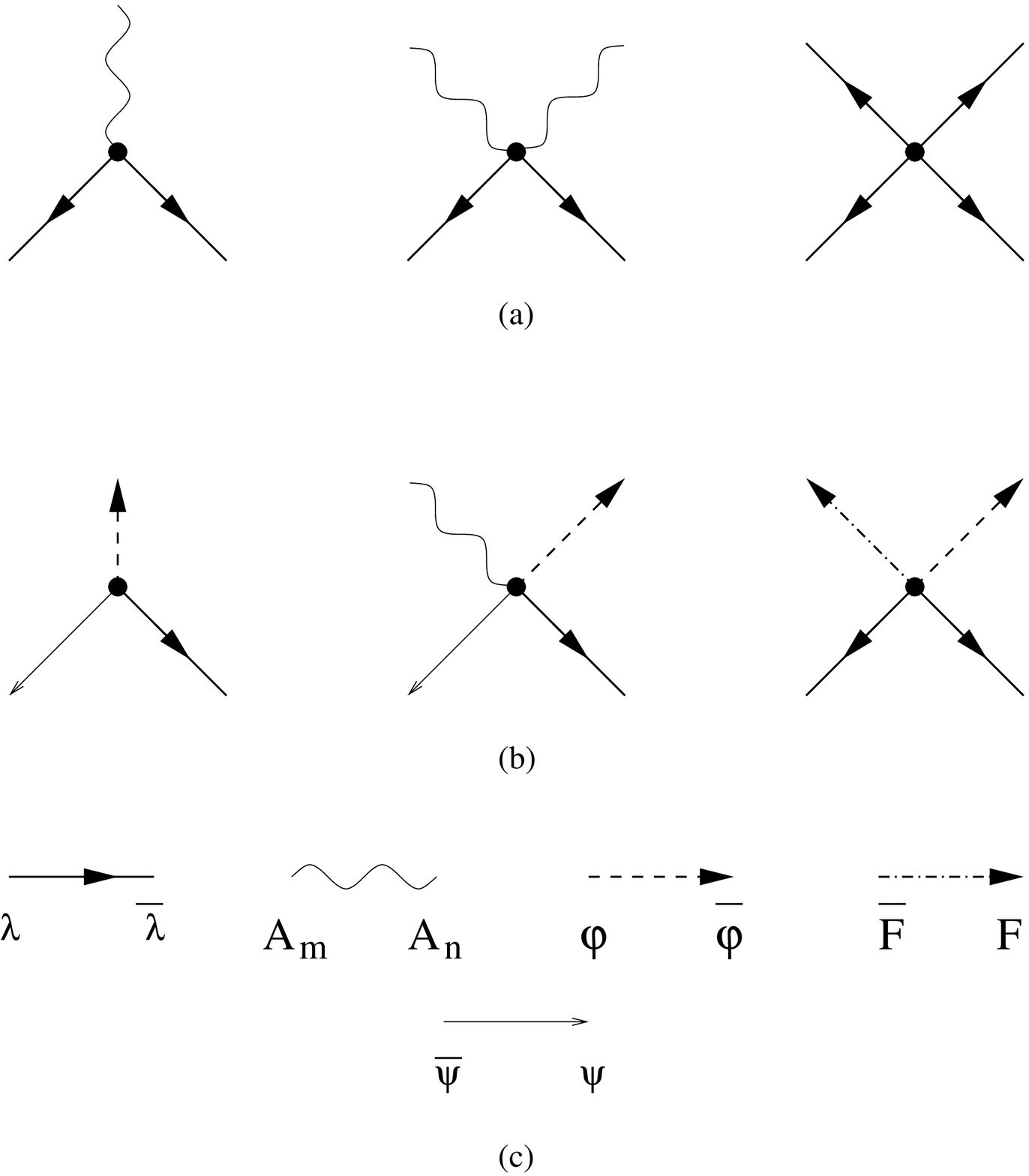,width=13cm} {\em Deformed part $L_{{\cal
N}=1/2}$ breaks the R-charge and the hermiticity. Vertices from
(\ref{SQCD}) break R-charge by 2 or 4 : (a) deformation of the
Yang-Mills coupling (first two terms in the last line of
(\ref{SQCD})); (b) deformation of the gauge--matter coupling
(second two terms in the last line of (\ref{SQCD}). The  R-charge
flow is indicated by bold arrows. Propagators of Yang-Mills and
matter component fields are summarized in (c). \label{figOne} }
%%%%%%%%%%%%%%%%%%%%%%%%%%%%%%%%%%%%%%%%%%%%%%%%%%%%%%%%%%%%%%%%%
\section{Proof of Renormalizability}
%%%%%%%%%%%%%%%%%%%%%%%%%%%%%%%%%%%%%%%%%%%%%%%%%%%%%%%%%%%%%%%%%
To show that the theory (\ref{SQCD}) is renormalizable, we
begin with the power-counting. Let us characterize $\ell$--th line
in a given Feynman loop diagram $L$ by two numbers: $r_\ell$ which
counts the power of momentum in the propagator ($r_\ell=-2$ for a
boson and $r_\ell=-1$ for a fermion) and $d_\ell$ which counts the
number of derivatives which act on a given propagator. For each
vertex labeled by $i \in L$, we introduce an index $\omega_i$:
\bea \omega_i=\frac{1}{2}\sum_{\ell \in i} (r_\ell+d_\ell+4)-4
\nonumber \eea
where sum is performed over the propagators coming to the chosen
vertex $i$. Then the `superficial degree of divergence' for the
loop diagram $L$ is given by
\bea\label{IndexStart} \Omega_{\rm div.}(L) =\sum_{i \in L}
\omega_i-\left( \frac{1}{2}\sum_{\rm ext}(r_\ell+d_\ell+4) -4
\right), \nonumber \eea
 where the last sum is performed only over the external lines.
 The Feynman loop diagram $L$ is superficially divergent if
 $\Omega_{\rm div}(L) \ge 0$.

We recall a quantum theory on $\mathbb{R}^4$ is referred
renormalizable if it contains only a finite number of `basic'
diagrams with $\Omega_{\rm div}\ge 0$. For ordinary (hermitian)
theories in $\mathbb{R}^4$ renormalizability leads to the
requirement that the theory contains interaction vertices with
$\omega_i\le 0$ only \footnote{This statement is equivalent to the
absence of operators with field-dimensions higher then four.}, and
contain only a finite number of `basic' diagrams with $\Omega_{\rm
div}\ge 0$. The first two lines of (\ref{SQCD}) describe such
renormalizable theory, and is in fact the ${\cal N}=1$
supersymmetric gauge theory. On the other hand, the last line of
(\ref{SQCD}) consist of operators with field-dimensions five or
higher, but we will now show that the theory still contains a
finite number of `skeleton' divergent diagrams and thus it is
renormalizable.

Before proceeding further, let us pause with recalling the
standard argument that a field theory containing vertices with
$\omega>0$ is necessarily nonrenormalizable, as the argument fails
for the class of theories under consideration. In the standard
argument, one {\it assumes} that starting from any (convergent)
loop diagram, one can add to it vertices with $\omega>0$, thus
making the diagram divergent. This assumption is justified by the
hermiticity. For example, one can always add a vertex with
$\omega> 0$ and its hermitian-conjugate. In a class of theories
under consideration, the hermiticity is lost, as is evident from
the observation that chiral Grassman-odd coordinates are deformed
but not antichiral ones. The lack of hermiticity invalidates the
conclusion drawn from the standard argument, and is ultimately
responsible for renormalizability of the theory.

Thus, for the proof of renormalizability of (\ref{SQCD}), we will
make extensive use of lack of hermiticity: many of the symmetries
present in the undeformed ${\cal N}=1$ theory are violated in the
deformed ${\cal N}={1 \over 2}$ theory. In keeping track of them,
the most useful one is the (pseudo) R-symmetry:
\bea \mbox{U}_{\rm R}(1): \qquad \A_f \rightarrow
e^{-i\alpha}\A_f, \qquad F_f \rightarrow e^{+i\alpha}F_f,\qquad
\la\rightarrow e^{-i\alpha}\la,\qquad C^{\alpha\beta}\rightarrow
e^{-2i\alpha}C^{\alpha\beta} \label{rsymm} \eea
and ${\O F}$, ${\bar \varphi}$ and ${\bar\la}$ transforming with
opposite charge and all the other fields being neutral. The
Lagrangian (\ref{SQCD}) is invariant under this pseudo R-symmetry,
and there is only one coupling constant $C^{\alpha\beta}$ which is
charged under it. The lack of hermiticity is reflected by the fact
that there is no $C^{\dagger {\dot\a}{\dot\b}}$ that can be
assigned with the opposite R-charge. Stated differently, all the
operators in $L_{{\cal N}=1/2}$ violate R-symmetry only by
positive value of R-charges. This means that any loop diagram
which contains vertices from the last line of (\ref{SQCD}) should
have enough external lines to accommodate the conservation of the
R-charge. As each vertex corresponding to an interaction with
coupling constant $C^{\a\b}$ or $|C|^2$ breaks the R {symmetry}
(but it still preserves the pseudo R-symmetry (\ref{rsymm})), so
there ought to be some lines carrying positive R-charge which
originate from such vertex. We illustrate this in fig.
\ref{figOne}. Since theory does not have vertices which decrease
the R charge (by the lack of hermiticity), one can trace the
``flow'' of the positive R-charge all the way to the external
lines. We refer the external states which lie on the `lines of
R-charge flow' as `R-charge violating states', and it is
convenient to separate them from the sum:
\bea -\frac{1}{2}\sum_{\rm
ext}(r_\ell+d_\ell+4)=-\frac{1}{2}\sum_{\rm R\
viol}(r_\ell+d_\ell+4) -\frac{1}{2}{\sum_{\rm
ext}}'(r_\ell+d_\ell+4)\le -N_R -\frac{1}{2}{\sum_{\rm
ext}}'(r_\ell+d_\ell+4). \nonumber \eea
Here $N_R$ is the number of the `lines of R-charge flow', and the
summation $\sum'$ does not include `R-charge violating lines'.
Since the `lines of R-charge flow' can be traced back to the
vertices from the third line of (\ref{SQCD}), for the superficial
degree of divergence (\ref{IndexStart}), we now get:
\bea \Omega_{\rm div}\le\sum
(\omega_i-N^{(i)}_R)+4-\frac{1}{2}{\sum_{\rm
ext}}'(r_\ell+d_\ell+4). \nonumber \eea
Here, $N_R^{(i)}$ denotes the R-charge violated by a given vertex.
We thus see that the index of the vertex $\omega_i$ is effectively
replaced by ${\hat\omega}_i=\omega_i-N_R^{(i)}$ and the theory
ought to be renormalizable if and only if this new index is
non--positive for all vertices.

For the vertices originating from the undeformed part, viz. the
ordinary ${\cal N}=1$ supersymmetric gauge theory, the new index
${\hat\omega}_i$ reduces to $\omega_i\le 0$. For the vertices
originating from the deformed part, viz. those ${\cal N}={1 \over
2}$ terms arising from non(anti)commutative deformation, we have:
\bea \mbox{Tr}\left\{\d_{[m}A_{n]}{\O \la}{\O \la}\right\},\,
\quad (\d_m {\O \A}_f){\O \la}^{\dot\alpha}\psi_{\beta \,
f},\,\quad
{\O \A}_f \d_{[m}A_{n]}F_f:
&&\quad \omega=\frac{1}{2},\quad N_R=2,\quad {\hat\omega}=
-\frac{3}{2}
\nonumber\\
\mbox{Tr}\left\{[A_{m}A_n]{\O \la}{\O \la}\right\},\quad (A_m {\O
\A}_f){\bar\la}^{\dot\alpha}\psi_{\beta \, f},
\,\quad
{\O \A}_f A_{[m}A_{n]}F_f:&&\quad
\omega=1,\quad N_R=2,\quad {\hat\omega}= -1\nonumber\\
\mbox{Tr}({\O \la}{\O \la})^2,\qquad\qquad {\O \A}_f {\bar\la}{\O
\la}F_f \quad :&&\quad \omega=2,\quad N_R=4,\quad {\hat\omega}=
-2. \nonumber \eea
We see that ${\hat\omega}\le 0$ for all of these vertices. We have
thus established the power-counting asserting that the
non(anti)commutative ${\cal N}={1 \over 2}$ supersymmetric gauge
theory is renormalizable. It now remains to classify all possible
operators and counterterms that are subject to renormalization.
%%%%%%%%%%%%%%%%%%%%%%%%%%%%%%%%%%%%%%%%%%%%%%%%%%%%%%%%%%%%%%%%%
\section{Operator Analysis}
%%%%%%%%%%%%%%%%%%%%%%%%%%%%%%%%%%%%%%%%%%%%%%%%%%%%%%%%%%%%%%%%%
To identify operators and counterterms needed for renormalization,
we will utilize operator analysis and various symmetry arguments
including the chiral flavor symmetry (\ref{flavor}) and the pseudo
R-symmetry (\ref{rsymm}). The most general, gauge-invariant, and
${\cal N}={1 \over 2}$ supersymmetric local operators that can
appear through radiative corrections are expressible
as\footnote{This expression should be understood as schematic: to
simplify notation, we write component fields of $\Phi_f$, but not
those of ${\tilde\Phi}_{\tilde f}$. However, as they carry the
same R-charge and mass-dimensions, the schematic notation does not
affect our argument. To be more explicit, one should take, for
example, $\phi_f^{\alpha_1}{\tilde\phi}_{\tilde
f}^{{\tilde\alpha}_1}$ instead of $\phi_f^{\alpha_1}$ and use
${\alpha}_1+{\tilde\alpha}_1$ wherever we have $\alpha_1$.}.
\bea \{ {\cal O}_\Lambda \} = \Lambda^\beta \, C_{mn}^\alpha \,
\d_m^{\alpha_0} \cdot \A_f^{\alpha_1}{\bar \A}_f^{{\bar\alpha}_1}
\, F_f^{\alpha_2}{\bar F}_f^{{\bar\alpha}_2}\,
\psi_f^{\alpha_3}{\bar\psi}_f^{{\bar\alpha}_3} \cdot
A^{\alpha_4}_m \la^{\alpha_5} {\bar \la}^{{\bar\alpha}_5} \,
D^{\alpha_6}, \nonumber \eea
where $\Lambda$ is an ultraviolet cutoff scale and $\alpha_i$'s
are non-negative integers. Notice that, apart from the
non(anti)commutativity parameter $C^{mn}$, there is no parameter
in the theory with mass-dimension. This operator (including the
coupling $C^{mn}$) should carry the mass-dimension four:
\bea\label{PwoCount1}
\beta-\alpha+\alpha_0+\alpha_1+{\bar\alpha}_1+\alpha_4+
2(\alpha_2+{\bar\alpha}_2+\alpha_6)+
\frac{3}{2}(\alpha_3+{\bar\alpha}_3+\alpha_5+{\bar\alpha}_5)=4
\eea
 and the pseudo R-charge zero:
 \bea\label{Rcharge}
-2\alpha+\alpha_2+{\bar\alpha}_1+{\bar\alpha}_5-{\bar\alpha}_2-
{\alpha}_1-{\alpha}_5=0. \eea
 It follows from the last expression that
 \bea\label{PwoCount2a}
{\bar\alpha}_5=2\alpha+{\bar\alpha}_2+{\alpha}_1+{\alpha}_5
-{\bar\alpha}_1-{\alpha}_2. \eea
Then, (\ref{PwoCount1}) yields an equation for the remaining
powers:
\bea\label{PwoCount2}
\beta+2\alpha+\alpha_0-\frac{1}{2}{\bar\alpha}_1+\frac{5}{2}{\alpha}_1+\alpha_4+
\frac{1}{2}\alpha_2+\frac{7}{2}{\bar\alpha}_2+2\alpha_6+
3\alpha_5+\frac{3}{2}(\alpha_3+{\bar\alpha}_3)=4. \eea
%

%%%%%%%%%%%%%%%%%%%%%%%%%%%%%%%%%%%%%%%%%%%%%%%%%%%%%%%%%%%%%%%
\subsection{Class I: Operators without $\O \varphi$ }
%%%%%%%%%%%%%%%%%%%%%%%%%%%%%%%%%%%%%%%%%%%%%%%%%%%%%%%%%%%%%%%

Equation (\ref{PwoCount2}) is useful for analyzing operators with
${\bar\alpha}_1=0$. We see that such operators have $\alpha\le 2$
and the only counterterm with $\alpha=2$ has a schematic form:
\bea (C_{mn})^2 \O \la \O \la \, \O \la \O \la. \nonumber \eea
Analyzing indices more carefully, we find four possible types of
operators with $\alpha=2$:
\bea {\cal O}_1={\mbox{det}}C~{\mbox{Tr}}( \O \la \O \la \, \O \la
\O \la), \quad&& {\cal O}_2= |C|^2~{\mbox{Tr}} ({\O \la}{\O
\sigma}_{mn}{\O\la}
{\O \la}{\O \sigma}^{mn}{\O \la}),\nonumber\\
{\cal O}_3=C_{mn}C_{pq}{\mbox{Tr}}({\O \la}{\O \sigma}^{mn}{\O
\la} {\O \la}{\O \sigma}^{pq}{\O \la}),\quad&& {\cal
O}_4=C_{mn}C_{pq}{\mbox{Tr}}({\O \la}{\O \sigma}^{mn}{\O \la} {\O
\la}{\O \sigma}^{pq}{\O \la}). \nonumber \eea
Using Fierz identities of four-fermion and self--duality of
$C_{mn}$, one can show that there is only one independent operator
${\cal O}_1$ and it is already present in the classical Lagrangian
(\ref{SQCD}).

Consider ${\bar\alpha}_1=0,\alpha=1$. In order to be able to
contract indices in ${\cal O}$, we should either have
$\alpha_0+\alpha_4\ge 2$, or we should be able to introduce
$\sigma$ matrices into the vertex. We consider these two cases
separately.

(a) For $\alpha_0+\alpha_4\ge 2$, (\ref{PwoCount2}) shows that the
only nonvanishing powers are $\alpha_0,\alpha_4$. We also have
${\bar\alpha}_5=2$ and $\alpha_0=2-\alpha_4$. So we have the
following candidates:
\bea C_{mn}\d^2 {\bar\la}{\bar\la},\qquad C_{mn}\d A_m
{\bar\la}{\bar\la},\qquad C_{mn} A_m A_n {\O \la}{\O \la}.
\nonumber \eea
 Using antisymmetry of $C_{mn}$ and
gauge invariance, we see that there is only one possible operator:
\bea C_{mn}F^{mn}{\bar\la}{\bar\la} \nonumber \eea
and this is the operator already present in the classical
Lagrangian (\ref{SQCD}).

(b) For vertices with $\sigma$ matrices, we should have at least
two fermions. From (\ref{PwoCount2}) we see that $\alpha_5=0$ and
we can not have more than one fermion. There are only the
following operators:
\bea C_{mn}{\O \psi}_f{\O \sigma}^{mn}{\O \la}F_f,\qquad C_{mn}{\O
{\tilde\psi}}_{\tilde f}{\O \sigma}^{mn}{\O \la} {\tilde
F}_{\tilde f},\qquad C_{mn}{\O \la}{\O \sigma}^{mn}{\O \la},
\nonumber \eea
 but they vanish identically by virtue of self--duality of $C^{mn}$.

%%%%%%%%%%%%%%%%%%%%%%%%%%%%%%%%%%%%%%%%%%%%%%%%%%%%%%%%%%%%%%%%%%
\subsection{Class II: Operators involving $\O \varphi$}
%%%%%%%%%%%%%%%%%%%%%%%%%%%%%%%%%%%%%%%%%%%%%%%%%%%%%%%%%%%%%%%%%%
For operators containing ${\O\varphi}$ it is more convenient to
extract ${\bar\alpha}_1$ from (\ref{Rcharge}), then instead of
(\ref{PwoCount2a}) and (\ref{PwoCount2}) we get:
\bea\label{PwoCount3a}
{\bar\alpha}_1={\bar\alpha}_2+\alpha_1+\alpha_5+2\alpha
-\alpha_2-{\bar\alpha}_5
\eea
and
\bea\label{PwoCount3}
\beta+\alpha+\alpha_0+\frac{1}{2}{\bar\alpha}_5+2{\alpha}_1+\alpha_4+
\alpha_2+3{\bar\alpha}_2+2\alpha_6+
\frac{5}{2}\alpha_5+\frac{3}{2}(\alpha_3+{\bar\alpha}_3)=4 \eea
From (\ref{PwoCount3}) we see that there are no operators with
$\alpha>4$. Let us analyze operators with different $\alpha\le 4$.

$\bullet$ For $\alpha=4$ we get only one possible operator:
\bea (C_{mn})^4 (\O \A_f \O {\wt \A}_{\tilde f})^4 \nonumber
\eea
but it is inconsistent with the ${\rm U}_{\rm A}(1)$ chiral flavor
symmetry.

$\bullet$ For $\alpha=3$ we need at least two indices from the
operator to contract with the parameter $C^{mn}$ (since
$C^{mn}C^{np}C^{pm}=0$), so we have two options:

(a) For $\alpha_0+\alpha_4\ge 2$, (\ref{PwoCount3}) has no
solution.

(b) With at least two fermions, solving (\ref{PwoCount3}), we get
an operator
\bea (C^{mn})^3 {\O \la \O \la} (\O \A_f \O {\wt
\A}_{\wt f})^2, \nonumber \eea
but it is not consistent with the ${\rm U}_{\rm A}(1)$ chiral
flavor symmetry.

$\bullet$ For $\alpha=2$, it is convenient to classify the
operators based on the number of fermions:

(a) For the operators with at least one fermion from the matter
multiplet, (\ref{PwoCount3}) gives only one operator:
\bea (C^{mn})^2({\O \A}_f \O {\wt \A}_f) \O \A_{f'}{\O
\la}\psi_{f'}, \nonumber \eea
but again it is not consistent with the ${\rm U}_{\rm A}(1)$
chiral flavor symmetry.

(b) Operators with four ${\O \la}$ have ${\O \alpha}_1=0$ and they
are analyzed already.

(c) Operators with two ${\O \la}$ must have $\alpha_2={\O
\alpha}_1$ to preserve the ${\rm U}_{\rm A}(1)$ chiral flavor
symmetry, so the only operator with nonzero ${\bar\alpha}_1$ is
\bea\label{ProblOper} (C^{mn})^2{\O \A}_f{\O
\la}{\O\la}F_f, \qquad (C^{mn})^2{\O {\wt \A}}_{\wt f}{\O
\la}{\O\la}\wt F_{\wt f}. \eea
This operator is present already in the classical Lagrangian
(\ref{SQCD}). There are also other operators:
\bea\label{ChirBreak} \Lambda (C^{mn})^2{\O \A}_f {\O \la}{\O
\la}{\O {\wt \A}}_{\wt f} ,\qquad (C^{mn})^2\d_k{\O
\A}_f{\bar\la}{\bar\la}{\O {\wt \A}}_{\wt f},\qquad
(C^{mn})^2A_k{\O \A}_f {\O \la}{\O \la} {\O {\wt \A}}_{\tilde f}.
\eea
These operators violate the chiral flavor ${\rm U}_{\rm A}(1)$
symmetry and hence cannot appear for massless matter, but, the
first operator with linear UV-divergence is potentially dangerous
operator for the massive matter. We will analyze this case
separately in the next subsection.

(d) For operators without fermions, we again have
$\alpha_2={\bar\alpha}_1$, so the only operators with
${\bar\alpha}_1\ne 0$ are
\bea (C^{mn})^2({\O \A}_f F_f)^2, \qquad
(C^{mn})^2({\O \A}_f F_f) (\O {\wt \A}_{\wt f} \wt F_{\wt f}),
\qquad (C^{mn})^2({\O {\wt \A}}_{\wt f} \wt F_{\wt f})^2 .
\eea
These operators are consistent with all symmetries of the Lagrangian, so
they will be generated on quantum level. This means that one should
introduce these terms even in the classical Lagrangian to make the theory
renormalizable.

$\bullet$~For $\alpha=1$ and nonzero ${\bar\alpha}_1$, to be able
to contract the indices of $C^{mn}$, we again should have one of
the following situations:

(a) For $\alpha_0+\alpha_4\ge 2$, (\ref{PwoCount3}) shows that
either $\alpha_0+\alpha_4=3$, or $\alpha_0+\alpha_4=2$,
$\alpha_2+\beta=1$. In both cases ${\bar\alpha}_1=2-\alpha_2$,
so the chiral symmetry (\ref{flavor}) requires
${\bar\alpha}_1=\alpha_2=1$. This allows only one term consistent with
gauge symmetry:
\bea
C^{mn}{\O \A}_f F_{mn}F_f
\eea
which was already present in the Lagrangian.

(b)~ For vertices with $\sigma$-matrices:
\bea\label{tryOne} C_{mn}{\O \A}_f\psi_{\alpha \,
f}(\sigma^{mn})_{\alpha\beta}{\O \A}_{f'} \psi_{\beta, f'},\quad
C_{mn}A_m{\O \A}_f{\O \la}\sigma^n\psi_f,\quad C_{mn}\d_m{\O \A}_f
{\O \la}\sigma^n\psi_f,\quad C_{mn}{\O \A}_f \la
\sigma{\O\la}\,{\O {\wt \A}}_{\wt f}. \eea
Notice that the first vertex vanishes due to the Fermi statistics,
and the last vertex cannot be contracted with $C_{mn}$ (there is
nothing like $(\sigma^{mn})_{\dot\alpha \beta}$). The other two
vertices in (\ref{tryOne}) have to combine into a gauge-invariant
operator
\bea C^{\alpha\beta}\sigma^m_{\alpha{\dot\alpha}} \nabla_m {\O
\A}_f{\bar\la}^{\dot\alpha}\psi_{\beta \, f}, \nonumber \eea
but this operator is present already in the classical Lagrangian
(\ref{SQCD}). Equations (\ref{PwoCount3a}), (\ref{PwoCount3}) have
a solution for another vertex involving $\sigma$-matrices
(${\bar\alpha}_5=2,{\bar\alpha}_1=0$), but we already analyzed
this type of operators above.

This concludes consideration of all possible operators with
$\alpha>0$, and the diagrams with $\alpha=0$ have only the
vertices of undeformed ${\cal N}=1$ supersymmetric gauge theory,
which is renormalizable by itself.

\subsection{Massive Matter}
Recall that the theory (\ref{SQCD}) contains both $\Phi_f$ and it
conjugate ${\wt \Phi}_{\wt f}$. For massless matter, there is no
coupling between $\Phi_f$ and ${\wt\Phi}_{\wt f}$, so Lagrangian
is symmetric under chirality transformations (\ref{flavor}). There
is also the ${\rm U}_{\rm F}(1)$ flavor symmetry:
\bea {\rm U}_{\rm F}(1): \qquad \Phi_f\rightarrow
e^{i\delta}\Phi_f,\qquad {\wt\Phi}_{\wt f}\rightarrow
e^{-i\delta}{\wt\Phi}_{\wt f}. \label{u1flavor} \eea
These two symmetries combined with the fact that all counterterms
have no more than one power of $\Phi_f$ and ${\wt\Phi}_{\wt f}$
(so that the operators like $({\O \Phi}_f \Phi_f)({\O
{\wt\Phi}_{\wt f}} {\wt\Phi}_{\wt f})$ do not appear) ensure that
the proof of renormalizability works separately for terms
involving $(\A,\psi,F)_f$ and $({\wt \A},{\wt\psi},{\wt F})_{\wt
f}$.

Our discussion so far was restricted to the case of massless
theory (\ref{SQCD}). However, the main result (absence of
counterterms which were not present in the original Lagrangian)
can be easily extended to the massive case as follows.

If mass is not equal to zero, then chiral symmetry is broken by
the mass parameters: \bea L_m=m\left\{\A_f{\wt F}_{\wt f}+{{\wt
\A}_{\wt f}}{F}_f-\psi_f {\wt \psi}_{\wt f}\right\}+ {\O
m}\left\{{\O \A}_f{\O {\wt F}}_{\wt f}+{\O{\wt \A}_{\wt f}}{\O
F}_{f}- {\O\psi}_{f} {\O{\wt\psi}_{\wt f}}\right\}. \eea
Even though we ultimately want to consider real mass $m$, in
${\cal N}={1 \over 2}$ theory, $m$ and ${\bar m}$ are independent
coupling parameters. As such, the ${\rm U}_{\rm A}(1)$ chiral
flavor symmetry (\ref{flavor}) is replaced for massive matter by a
pseudo ${\rm U}_{\rm A}(1)$ chiral-symmetry:
\bea {\rm U}_{\rm A}(1): \qquad \Phi\rightarrow
e^{+i\gamma}\Phi,\qquad {\tilde\Phi}\rightarrow
e^{+i\gamma}{\tilde\Phi},\qquad m\rightarrow e^{-2i\gamma}m .
\label{newchiralu1} \eea
All possible counterterms in the theory must be invariant under
the new pseudo ${\rm U}_{\rm A}(1)$ chiral-symmetry.

In our discussion of massless case, we classified explicitly all
possible operators and counterterms which were consistent with the
pseudo R-symmetry, but not necessarily with the chiral flavor
symmetry. We then argued that some of the terms were ruled out by
the chiral symmetry. To discuss the massive theory, we just have
to reconsider those operators and see whether some of them which
were prohibited before by the chiral flavor symmetry might be
consistent with the new pseudo ${\rm U}_{\rm A}(1)$
chiral-symmetry.

\EPSFIGURE[ht]{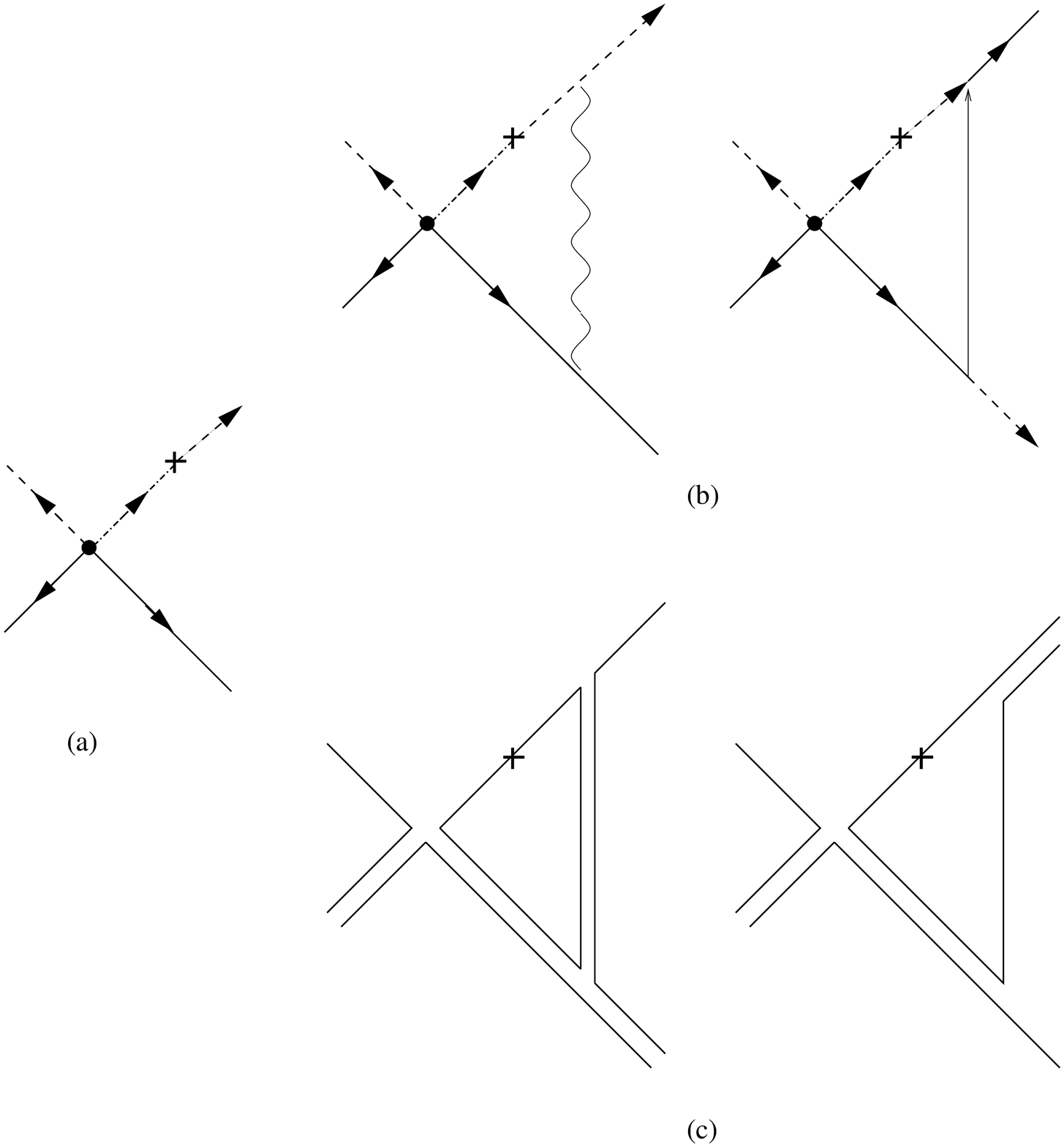,height=13cm,width=13cm} {\sl New vertex
(\ref{NewVertex}) arising in massive matter theory: (a) tree-level
vertex, (b) one loop contribution to renormalization of the
tree-level vertex in the usual notation; (c) the same diagrams in
the double line notation. We see that one loop diagrams contribute
at different orders in $1/N$. Cross indicates mass
insertion.\label{figTwo} }

We will use the chiral perturbation theory (i.e. perturbation theory
in $m$ and ${\bar m}$). Let us discuss various cases separately.

(a) If operator does not contain any power of $m$ or ${\O m}$,
then it is invariant under chiral {\it symmetry} and such
operators have been analyzed above.

(b) In the first order in $m$ or ${\O m}$, we have operators which
break the chiral flavor symmetry by two units. On the other hand,
since $m, \O m$ carry dimension one, such operator can arise only
from the one which diverges linearly with $\Lambda$, viz. by
substituting $\Lambda$ by $\O m$. We had only one such operator:
first term of (\ref{ChirBreak}), so we have two candidates:
\bea\label{NewVertex} {\O m} |C|^2{\O {\wt \A}}_{\wt
f}{\O\la}{\O\la}{\O \A}_f,\qquad {\O m}
|C|^2{\O{\A}}_f{\O\la}{\O\la}{\O {\wt \A}_{\wt f}}. \eea
These two operators are consistent with all symmetries of the
Lagrangian. In fact, they arise even at tree level in the chiral
perturbation theory (see figure 2a). The tree-level vertex is
radiatively corrected by the vertex renormalization in figure 2a
as well as the mass renormalization. Moreover, there are
additional renormalizations of the newly generated tree-level
vertex. The one-loop diagrams which would contribute to such
renormalization are presented in figure 2b. Each of these two
diagrams yields a nonzero contribution to the renormalization of
the vertex. The loop integral is the same in magnitude and
opposite in sign, but the two cannot cancel each other as they
have different orders in the color factor, $N$, as illustrated by
`t Hooft double-line notation in figure 2c.

(c) To have operators with higher power of $m$ and ${\bar m}$, we
have to start from a counterterm which diverges as $\sim
\Lambda^s$ with $s>1$ in the massless theory. We found that the
counterterms grows at most linearly with $\Lambda$ (disregarding
violation of the chiral flavor symmetry), so no operator with
powerlike UV divergences are generated beyond first order in the
chiral perturbation theory.

To summarize, we have analyzed all possible counterterms
consistent with symmetries of deformed ${\cal N}={1 \over 2}$
supersymmetric gauge theory, and we have shown that this theory is
renormalizable by itself, viz. without introducing any new
operators.
%%%%%%%%%%%%%%%%%%%%%%%%%%%%%%%%%%%%%%%%%%%%%%%%%%%%%%%%%%%%%%%%%%
\section{Star Wars: Confronting Renormalizability and Supersymmetry}
%%%%%%%%%%%%%%%%%%%%%%%%%%%%%%%%%%%%%%%%%%%%%%%%%%%%%%%%%%%%%%%%%%
So far, our proof for renormalizability was built largely on
various symmetries in the theory, but not much directly on the
${\cal N}={1 \over 2}$ supersymmetry or the non(anti)commutative
star product. In this section, we elucidate relation between them
and expose several intriguing features implicit to the operator
analysis in the previous section.

$\bullet$ In the foregoing analysis, we have taken the
noncommutativity parameter $\Theta^{mn}$ of the deformed
superspace to zero. If we turn on $\Theta^{mn}$ nonzero and render
the Grassman-even coordinates $y^m$ noncommutative as well, an
important modification arises regarding the radiative corrections.
In noncommutative field theories -- theories defined on spacetime
with $\Theta^{mn}$, an important feature of these theories
was the UV-IR mixing \cite{uvir}. Because of the noncommutativity,
fields are representable as `t Hooft double lines and Feynman
diagrams are classifiable into planar and nonplanar diagrams. The
star product phase-factors all cancel out for planar diagrams but
mixes external and loop momenta for nonplanar ones. Thus, any
explicit noncommutativity parameter dependence would arise from
nonplanar diagrams only. Put together, this leads to an important
consequence that nonplanar diagrams are free from UV divergences
(provided one keeps the external momenta nonzero).

From (\ref{star}), it is evident that the Grassman-even
phase-factors are correlated with the Grassman-odd phase-factors.
Therefore, planar or nonplanar diagrams classified according to
the Grassman-even phase-factors would be the same as those with
respect to the Grassman-odd phase-factors. This observation was
made explicit for non(anti)commutative Wess-Zumino model in
\cite{brittofengrey2}. Now, combined with the consequence of UV-IR
mixing for noncommutative field theories, this implies that terms
depending explicitly on $C^{\a\b}$ cannot have UV divergences, as
they originate only from nonplanar diagrams --- the otherwise UV
divergences are transmuted via UV-IR mixing to IR divergences. We
thus conclude that, in case $\Theta^{ab}$ nonzero, the theory is
renormalized only through planar diagrams.

$\bullet$ Our proof of renormalizability does not depend on the
${\cal N}={1 \over 2}$ supersymmetry. Notice that each of the
three terms proportional to $C^{\a\b}$ in the deformed Lagrangian
(\ref{SQCD}):
\bea {1 \over g^2} C^{mn} {\rm Tr} \{ F_{mn} \O \lambda \O \lambda
\}, \qquad C^{\a\b} \nabla_m \O \A_f \sigma^m_{\a \da} \O
\lambda^{\da} \psi_{\b \, f} \qquad  \mbox{and} \qquad C^{mn} \O
\varphi_f F_{mn} F_f \label{c1} \eea
are related under supersymmetry transformations
(\ref{susytransfv}), (\ref{susytransfm}) to the gaugino kinetic term
and the matter $\O F F$ term, respectively, but not to any other
terms. Therefore, numerical coefficients in front of the terms
(\ref{c1}) can be set arbitrary {\sl provided} numerical
coefficients of $C^{\a\b}$-dependent terms in the supersymmetry
transformations (\ref{susytransfv}), (\ref{susytransfm}) are
adjusted accordingly. Notice also that each of the two terms
proportional to $|C|^2$ in the deformed Lagrangian (\ref{SQCD}):
\bea (C^{mn})^2 {\rm Tr} (\O \lambda \O \lambda \, \O \lambda \O
\lambda) \qquad \mbox{and} \qquad (C^{mn})^2 \O \A_f ( \O \lambda
\O \lambda) F_f, \label{c2} \eea
is invariant under the ${\cal N}={1 \over 2}$ supersymmetry
transformations (\ref{susytransfv}), (\ref{susytransfm}) by
itself. This indicates that the ${\cal N}={1 \over 2}$
supersymmetry alone would not fix the numerical coefficients of
these terms and hence can be taken arbitrary values.

Our proof of renormalizability still holds without modification
even if the numerical factors in front of these terms were chosen
different from the ones in the Lagrangian.

$\bullet$ In the Lagrangian (\ref{SQCD}), numerical coefficients
in front of the terms (\ref{c1}) and (\ref{c2}) are determined by
the algebraic structure the star product (\ref{star}). The two
coefficients of terms (\ref{c1}) are correlated with the two
coefficients of $C^{\a\b}$-dependent supersymmetry transformations
in (\ref{susytransfv}), (\ref{susytransfm}). They are thus fixable
by normalization of $C^{\a\b}$ in the definition of the star
product (\ref{star}). Then, the remaining two coefficients of
terms (\ref{c2}) would reflect the algebraic structure of the star
product, since they arise from second order upon expanding the
exponential in (\ref{star}). It implies that the star product
representation for non(anti)commutativity is a structure imposed
in addition to the ${\cal N}={1 \over 2}$ supersymmetry. Notice
that the non(anti)commutativity deformation in (\ref{nac}):
\bea \th^\a \star \th^\b + \th^\b \star \th^\a = C^{\a\b},
\nonumber \eea
which leads immediately to the exponential form of the star
product, is a structure a priori independent of supersymmetry.

The above consideration brings out the issue whether the star
product (\ref{star}) is stable against radiative corrections. The
operator analysis in section 4 indicates that renormalization
effects do not appear to relate or constrain size of radiative
corrections to the deformation terms (\ref{c1}), (\ref{c2}). The
renormalization factor $Z_C$ for the non(anti)commutativity
parameter $C^{\a\b}$ (multiplied by the renormalization scale
$\mu$) is computable from radiative corrections to the terms
(\ref{c1}). Similarly, the renormalization factor $Z_{C^2}$ for
the {\sl square} of non(anti)commutativity parameter $|C|^2$
(multiplied by $\mu^2$) is computable from radiative corrections
to the terms (\ref{c2}). In general, these renormalization factors
are related each other. For simplicity, consider the pure gauge
theory with no matter. If the theory were to retain exponential
structure of the star product (\ref{star}), the two
renormalization factors are constrained to be
\bea (Z_C)^2 = Z_{C^2}. \label{wardid} \eea
Though extremely interesting, in this work, we do not dwell on
computations testing the relation (\ref{wardid}), as dictated by
the star product structure. We remark, however, that
nonrenormalization of the star product is physically required,
else the Grassman-even coordinates carry noncanonical dimension at
quantum level. The nonrenormalizatoin is also necessary for the
emergence of open Wilson lines. From the computation of effective
superpotential in non(anti)commutative Wess-Zumino model, we
expect that open Wilson lines again play important roles in
non(anti)commutative gauge theories. Combined with gauge
invariance, it is suggestive that the relation (\ref{wardid})
would hold to all orders in perturbation theory.
%%%%%%%%%%%%%%%%%%%%%%%%%%%%%%%%%%%%%%%%%%%%%%%%%%%%%%%%%%%%%%%%%%
\section{Discussions}
%%%%%%%%%%%%%%%%%%%%%%%%%%%%%%%%%%%%%%%%%%%%%%%%%%%%%%%%%%%%%%%%%%
In this work, we have proven that the ${\cal N}={1 \over 2}$
supersymmetric gauge theory is renormalizable to all orders in
perturbation theory. Our proof is based largely on operator
analysis and symmetry arguments. As such, it brings in many
interesting questions worthy of further investigation.

$\bullet$ In proving the renormalizability, we have taken trivial
superpotential for the matter superfields, either $W = \O W = 0$
for massless case or $W = m \Phi_f \wt \Phi_{\wt f}, \O W = \O m
\O \Phi_f \O {\wt \Phi}_{\wt f}$ for massive case. By combining
the result of the present work and that of the Wess-Zumino model
\cite{grisaru}, we conjecture that non(anti)commutative Yang-Mills
theory coupled to arbitrary matter contents and superpotential
involving the most general, gauge-invariant quadratic and cubic
interactions is renormalizable to all orders in perturbation
theory. In this case, as in the case of the gauge theory with
massive matter (discussed in the previous section) and of the
Wess-Zumino model \cite{brittofengrey1, jungtay, grisaru},
operators that are not prohibited by underlying symmetry needs to
be added. Part of our conjecture asserts that there are only
finitely many such operators and that the newly added operators do
not ruin renormalizability.

$\bullet$ The deformed gauge theory consists of two parts:
ordinary part $L_{{\cal N}=1}$ with ${\cal N}=1$ supersymmetry,
and deformed part $L_{{\cal N}=1/2}$ with ${\cal N}={1 \over 2}$
supersymmetry. We have shown that the ordinary part $L_{{\cal
N}=1}$ receives radiative corrections only from itself and none
from the deformed part $L_{{\cal N}=1/2}$. This implies that
ordinary part of the Wilsonian effective action is renormalized
only at one-loop order. We also have shown that radiative
corrections to the deformation part $L_{{\cal N}=1/2}$ arise only
through interactions governed by the $L_{{\cal N}=1}$ part.
Underlying to the renormalizability was violation of R-symmetry
and helicity conservation and non-Hermiticity of $L_{{\cal
N}=1/2}$. It suggests that deformed part of the Wilsonian
effective action is renormalized at all orders in perturbation
theory, as can be inferred from the operator relations concerning
divergence of R-symmetry current, which is violated by the
deformed part $L_{{\cal N}={1 / 2}}$ as well as the anomaly. It
would be interesting to analyze in detail renormalization group
flow of the Wilsonian effective action.

\section*{Acknowledgement} We acknowledge R. Britto, B. Feng,
J. Maldacena and N. Seiberg for enlightening and useful
discussions. SJR was a Member at the Institute for Advanced Study
during this work. He thanks the School of Natural Sciences for
hospitality and for the grant-in-aid from the Fund for Natural
Sciences.

\end{document}